\documentclass[twocolumn,aps,showpacs]{revtex4}
\usepackage{epsfig}
\usepackage{lineno}
\begin{document}

\title{First measurement of proton-induced low-momentum dielectron radiation off cold nuclear matter}

\author{G.~Agakishiev$^{6}$, A.~Balanda$^{3}$, D.~Belver$^{18}$, A.~Belyaev$^{6}$,
J.C.~Berger-Chen$^{8}$, A.~Blanco$^{2}$, M.~B\"{o}hmer$^{9}$, J.~L.~Boyard$^{16}$, P.~Cabanelas$^{18}$,
S.~Chernenko$^{6}$, A.~Dybczak$^{3}$, E.~Epple$^{8}$, L.~Fabbietti$^{8}$, O.~Fateev$^{6}$,
P.~Finocchiaro$^{1}$, P.~Fonte$^{2,b}$, J.~Friese$^{9}$, I.~Fr\"{o}hlich$^{7}$, T.~Galatyuk$^{7,c}$,
J.~A.~Garz\'{o}n$^{18}$, R.~Gernh\"{a}user$^{9}$, K.~G\"{o}bel$^{7}$, M.~Golubeva$^{12}$, D.~Gonz\'{a}lez-D\'{\i}az$^{d}$,
F.~Guber$^{12}$, M.~Gumberidze$^{16}$, T.~Heinz$^{4}$, T.~Hennino$^{16}$, R.~Holzmann$^{4}$,
A.~Ierusalimov$^{6}$, I.~Iori$^{11,f}$, A.~Ivashkin$^{12}$, M.~Jurkovic$^{9}$, B.~K\"{a}mpfer$^{5,e}$,
T.~Karavicheva$^{12}$, I.~Koenig$^{4}$, W.~Koenig$^{4}$, B.~W.~Kolb$^{4}$, G.~Kornakov$^{18}$,
R.~Kotte$^{5}$, A.~Kr\'{a}sa$^{17}$, F.~Krizek$^{17}$, R.~Kr\"{u}cken$^{9}$, H.~Kuc$^{3,16}$,
W.~K\"{u}hn$^{10}$, A.~Kugler$^{17}$, A.~Kurepin$^{12}$, V.~Ladygin$^{6}$, R.~Lalik$^{8}$,
S.~Lang$^{4}$, K.~Lapidus$^{8}$, A.~Lebedev$^{13}$, T.~Liu$^{16}$, L.~Lopes$^{2}$,
M.~Lorenz$^{7,\ast}$, L.~Maier$^{9}$, A.~Mangiarotti$^{2}$, J.~Markert$^{7}$, V.~Metag$^{10}$,
B.~Michalska$^{3}$, J.~Michel$^{7}$, D.~Mishra$^{10}$, C.~M\"{u}ntz$^{7}$, L.~Naumann$^{5}$, Y.~C.~Pachmayer$^{7}$,
M.~Palka$^{3}$, Y.~Parpottas$^{15,14}$, V.~Pechenov$^{4}$, O.~Pechenova$^{7}$, J.~Pietraszko$^{7}$,
W.~Przygoda$^{3}$, B.~Ramstein$^{16}$, A.~Reshetin$^{12}$, A.~Rustamov$^{7}$, A.~Sadovsky$^{12}$,
P.~Salabura$^{3}$, A.~Schmah$^{a}$, E.~Schwab$^{4}$, J.~Siebenson$^{8}$, Yu.G.~Sobolev$^{17}$,
S.~Spataro$^{g}$, B.~Spruck$^{10}$, H.~Str\"{o}bele$^{7}$, J.~Stroth$^{7,4}$, C.~Sturm$^{4}$,
A.~Tarantola$^{7}$, K.~Teilab$^{7}$, P.~Tlusty$^{17}$, M.~Traxler$^{4}$, R.~Trebacz$^{3}$,
H.~Tsertos$^{15}$, T.~~Vasiliev$^{6}$, V.~Wagner$^{17}$, M.~Weber$^{9,h,\ast}$, C.~Wendisch$^{5}$,
J.~W\"{u}stenfeld$^{5}$, S.~Yurevich$^{4}$, Y.~Zanevsky$^{6}$}

\affiliation{
(HADES collaboration) \\\mbox{$^{1}$Istituto Nazionale di Fisica Nucleare - Laboratori Nazionali del Sud, 95125~Catania, Italy}\\
\mbox{$^{2}$LIP-Laborat\'{o}rio de Instrumenta\c{c}\~{a}o e F\'{\i}sica Experimental de Part\'{\i}culas , 3004-516~Coimbra, Portugal}\\
\mbox{$^{3}$Smoluchowski Institute of Physics, Jagiellonian University of Cracow, 30-059~Krak\'{o}w, Poland}\\
\mbox{$^{4}$GSI Helmholtzzentrum f\"{u}r Schwerionenforschung GmbH, 64291~Darmstadt, Germany}\\
\mbox{$^{5}$Institut f\"{u}r Strahlenphysik, Helmholtz-Zentrum Dresden-Rossendorf, 01314~Dresden, Germany}\\
\mbox{$^{6}$Joint Institute of Nuclear Research, 141980~Dubna, Russia}\\
\mbox{$^{7}$Institut f\"{u}r Kernphysik, Goethe-Universit\"{a}t, 60438 ~Frankfurt, Germany}\\
\mbox{$^{8}$Excellence Cluster 'Origin and Structure of the Universe' , 85748~Garching, Germany}\\
\mbox{$^{9}$Physik Department E12, Technische Universit\"{a}t M\"{u}nchen, 85748~Garching, Germany}\\
\mbox{$^{10}$II.Physikalisches Institut, Justus Liebig Universit\"{a}t Giessen, 35392~Giessen, Germany}\\
\mbox{$^{11}$Istituto Nazionale di Fisica Nucleare, Sezione di Milano, 20133~Milano, Italy}\\
\mbox{$^{12}$Institute for Nuclear Research, Russian Academy of Science, 117312~Moscow, Russia}\\
\mbox{$^{13}$Institute of Theoretical and Experimental Physics, 117218~Moscow, Russia}\\
\mbox{$^{14}$Frederick University, 1036~Nicosia, Cyprus}\\
\mbox{$^{15}$Department of Physics, University of Cyprus, 1678~Nicosia, Cyprus}\\
\mbox{$^{16}$Institut de Physique Nucl\'{e}aire (UMR 8608), CNRS/IN2P3 - Universit\'{e} Paris Sud, F-91406~Orsay Cedex, France}\\
\mbox{$^{17}$Nuclear Physics Institute, Academy of Sciences of Czech Republic, 25068~Rez, Czech Republic}\\
\mbox{$^{18}$LabCAF. Dpto. F\'{\i}sica de Part\'{\i}culas, Univ. de Santiago de Compostela, 15706~Santiago de Compostela, Spain}\\
\\
\mbox{$^{a}$ also at Lawrence Berkeley National Laboratory, ~Berkeley, USA}\\
\mbox{$^{b}$ also at ISEC Coimbra, ~Coimbra, Portugal}\\
\mbox{$^{c}$ also at ExtreMe Matter Institute EMMI, 64291~Darmstadt, Germany}\\
\mbox{$^{d}$ also at Technische Universit\"{a}t Darmstadt, ~Darmstadt, Germany}\\
\mbox{$^{e}$ also at Technische Universit\"{a}t Dresden, 01062~Dresden, Germany}\\
\mbox{$^{f}$ also at Dipartimento di Fisica, Universit\`{a} di Milano, 20133~Milano, Italy}\\
\mbox{$^{g}$ also at Dipartimento di Fisica Generale and INFN, Universit\`{a} di Torino, 10125~Torino, Italy}\\
\mbox{$^{h}$ also at University of Houston, Houston Texas, USA }\\
\\
\mbox{$^{\ast}$ corresponding authors: Lorenz@Physik.uni-frankfurt.de, m.weber@cern.ch}
}

\date{Received: 02.08.2012}

\begin{abstract}
We present data on dielectron emission in proton induced reactions on a Nb target at 3.5 GeV kinetic beam energy measured with HADES installed at GSI. The data represent the first high statistics measurement of proton-induced dielectron radiation from cold nuclear matter in a kinematic regime, where strong medium effects are expected. Combined with the good mass resolution of 2\%, it is the first measurement sensitive to changes of the spectral functions of vector mesons, as predicted by models for hadrons at rest or small relative momenta. Comparing the $e^+$$e^-$ invariant mass spectra to elementary p+p data, we observe for $e^+$$e^-$ momenta $P_{ee}<0.8$ GeV/c a strong modification of the shape of the spectrum, which we attribute to an additional $\rho$-like contribution and a decrease of $\omega$ yield. These opposite trends are tentatively interpreted as a strong coupling of the $\rho$ meson to baryonic resonances and an absorption of the $\omega$ meson, which are two aspects of in-medium modification of vector mesons.
\end{abstract}

\maketitle
\begin{figure}[htb]
   \mbox{\epsfig{figure={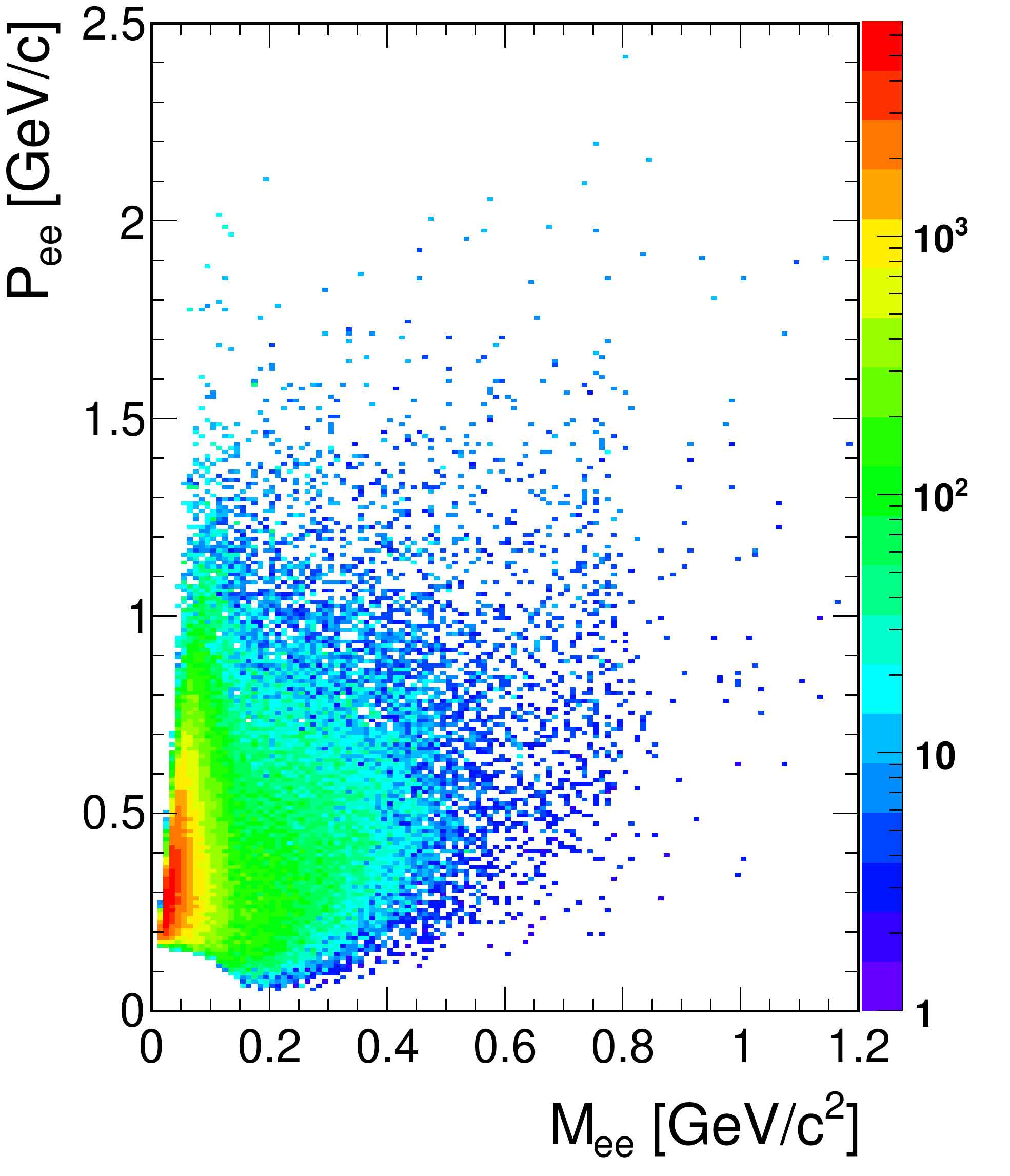}, width=0.59\linewidth}}
   \caption[]{Scatter plot of count rates of dielectrons as a function of invariant mass and momentum.}
\label{p_mass}
\end{figure}

The QCD vacuum is characterized by nonzero expectation values of various quark
and gluon operators. Most notable is the finite chiral quark condensate,
signalling the spontaneous breaking of chiral symmetry. In a cold strongly
interacting medium, the chiral condensate is expected to be modified (in
linear density approximation, by about 30\% at nuclear saturation density).
As predicted by various models \cite{weise,rho,lee,meissner,rapp,Leupold,Hatsuda,Friman,rapp1}, vector meson properties
should be affected by changes of the condensates. Such phenomena can be
studied in leptonic decays of the vector mesons  (e.g.\ $V \to e^+ e^-$, with
$V = \rho, \omega, \phi$). According to \cite{Hilger} the $\rho$ meson is
especially sensitive to changes of those condensates which break chiral symmetry (also called chirally odd condensates).
In hadronic many-body approaches many hints have been found for a broadening
of the $\rho$ meson spectral function in an ambient nuclear medium
\cite{rapp,Leupold,post,riek}.\\
Experimentally, in-medium properties can be studied in heavy-ion collisions (probing hot and dense hadronic matter) or in proton-, pion- or photon- induced reactions on nuclei (probing cold nuclear matter). For recent reviews on medium effects in cold nuclear matter see in particular \cite{Leupold,Hatsuda}. Medium modifications are expected to be stronger in heavy-ion collisions due to the higher densities and temperatures. Measured observables represent an average over the complete space-time evolution of the temperature and the density of the system. On the other hand, in proton-, pion- or photon- induced reactions the system does not undergo a noticeable density and temperature evolution in time hence conditions of the system are better defined.\\
Electron pair (e$^{+}$e$^{-}$) decay of vector mesons is an ideal probe for such studies since electrons and positrons are not affected by strong final state interactions.
A promising observable is the spectral shape, i.e. the e$^{+}$e$^{-}$ invariant mass distribution, but also the nuclear modification of the cross section provides valuable information since it is connected to the total widths of the hadrons inside the medium. The latter reflect also the coupling of mesons to resonance hole states, which is an important element of the in-medium self energy of propagating mesons.
However, two competing mechanisms have to be considered when discussing the nuclear modifications. Multi-step production mechanisms can enhance the particle production, while absorption of the produced hadrons reduces the detectable yields. Both contributions have to be considered before drawing solid conclusions about the hadron widths inside the medium.
A measurement sensitive to the spectral shape on the other hand requires that the decay takes place inside the nucleus. Therefore good acceptance for decays of low momentum vector mesons is of crucial importance, in particular for the relatively long living $\omega$ and $\phi$ mesons.\\
In fact all measurements focusing on the spectral distribution of dielectrons produced off nuclei in photon and proton induced reactions are restricted to relatively high momenta ($P_{ee}>0.8$ GeV/c) and are not conclusive yet. For the $\rho$ meson, the CLAS experiment at JLab \cite{clas} reports a slight broadening and no shift of the $\rho$ pole mass in photon induced reactions, while the E325 experiment at KEK \cite{kek} deduced a shift but no broadening in proton induced reactions.
For the $\omega$ and $\phi$ mesons, experiments \cite{cbelsa,spring,anke,clas2} report a sizable collisional broadening (up to a factor 16 larger than the natural line width in case of the $\omega$) inside the medium extracted by comparing the nuclear modification of the cross sections to microscopic transport models.
Besides the complications due to additional yield fed by pion driven secondary collisions inside the nucleus, the disadvantage of these indirect measurements is their model dependence. For instance, the recent analysis of \cite{rodrigues} led to a significantly different width of the $\omega$ meson inside the medium as extracted in \cite{cbelsa}.\\
In this letter we report on inclusive e$^{+}$e$^{-}$ pair production in proton induced reactions at E$_{kin} = 3.5$~GeV on the nucleus Nb, representing the first high statistics measurement with small momenta of e$^{+}$e$^{-}$ pairs relative to the medium ($P_{ee}<0.8$ GeV/c), see Fig.~\ref{p_mass}. By comparing the p+Nb data to e$^{+}$e$^{-}$ production from elementary p+p collisions at the same kinetic beam energy \cite{hades-pp}, the presented results are sensitive to both above discussed observables, namely line shape modifications and nuclear suppression of the production cross section.\\

The experiments were performed with the High Acceptance Dielectron Spectrometer (HADES) \cite{hades-nim} located at the GSI Helmholtzzentrum f\"ur Schwerionenforschung in Darmstadt, Germany.
In both experiments a proton beam with an intensity of $\sim10^7$ particles/s was used. From the p+p measurement \cite{hades-pp}, the inclusive production cross sections of all relevant e$^+$ e$^-$ pair sources ($\pi^0, \eta, \Delta, \omega$ and $\rho$) were deduced. The experiment with Nb was carried out with a 12-fold segmented Nb target. For details about the experimental setup, particle identification and the data analysis we refer to \cite{hades-pp}.\\
\begin{figure}[htb]
   \mbox{\epsfig{figure={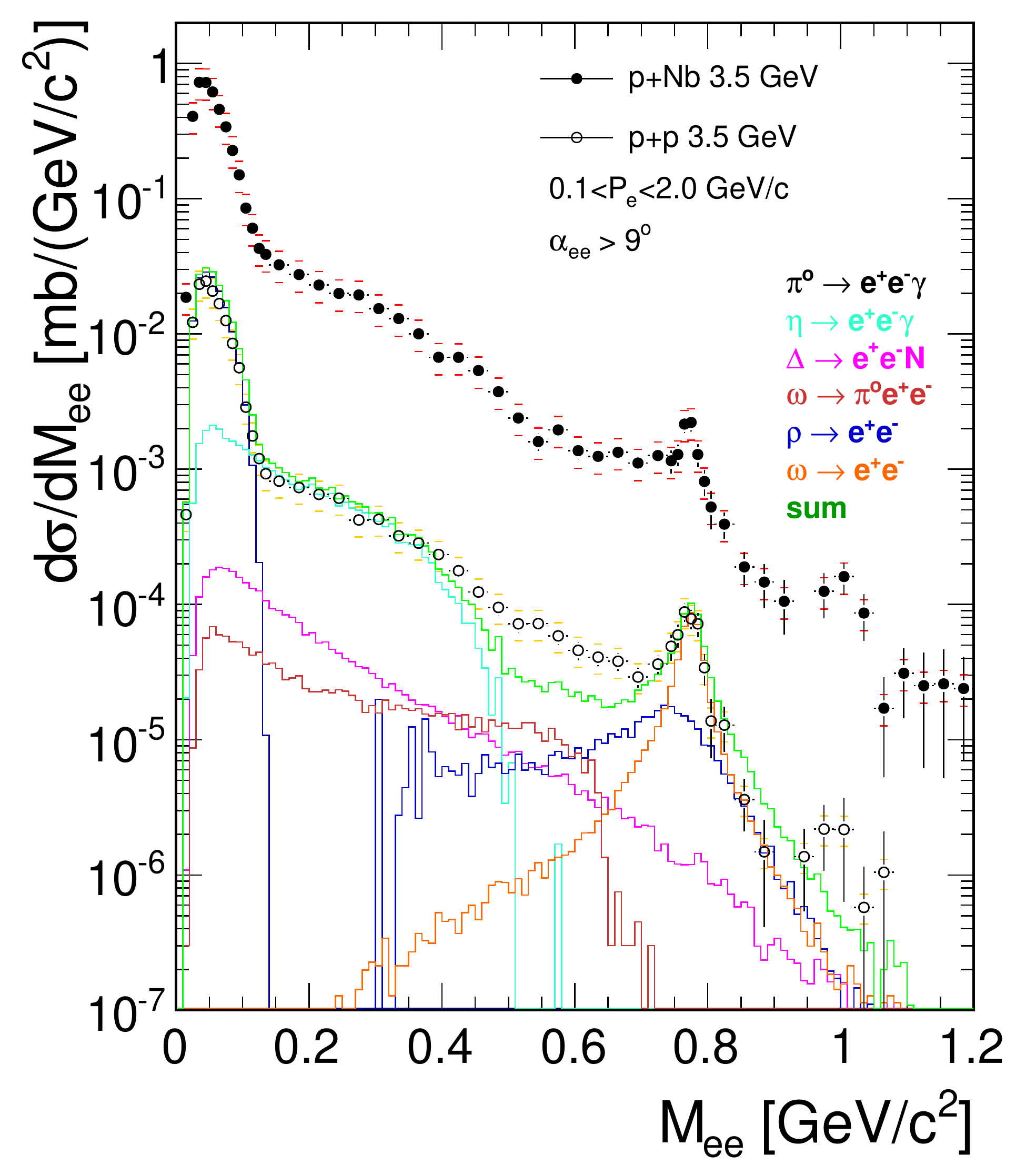}, width=0.9\linewidth}}
   \caption[]{Comparison of dielectron cross sections as a function of the invariant mass measured in p+p and p+Nb collisions. The p+Nb data are displayed with full circles and red horizontal lines indicating the systematical errors, while the p+p data are displayed with open circles and yellow horizontal lines. For the p+p data a PYTHIA dilepton cocktail is displayed in addition.}
\label{mass}
\end{figure}
Normalization of the dielectron cross sections have been obtained by measuring charged pions and by interpolating known pion production cross sections \cite{harp}.
The corresponding total reaction cross section amounts to  $\sigma_{pNb}=848\pm127$~[mb].
A detailed description of the procedure is given in \cite{pavel}.
In addition, the trigger efficiency of the first-level trigger, asking for a charged particle multiplicity larger than three, on inclusive e$^+$e$^-$ pair production $F$(e$^+$e$^-$)$=0.92$ was extracted from simulations and has been taken into account for the normalization of the dielectron distributions of the p+Nb run. In p+p collisions the normalization was obtained via the exclusive measurement of elastic p+p collisions and the known integrated cross section inside the HADES acceptance \cite{hades-pp}.\\
The efficiency corrected invariant mass distributions of e$^+$e$^-$ pairs are shown in Fig.~\ref{mass} for both collision systems.
The colored horizontal bars represent the systematic uncertainties, which result from the quadratic sum of errors estimated from the different particle identification methods (10\%), from consistency checks of the efficiency correction (10\%), including the uncertainty due to combinatorial background subtraction as well as the uncertainty from the normalization (15\%). The total systematic error amounts to 21\% in case of the p+Nb data while for the p+p data the systematic uncertainty is 20\%. For the comparison of the spectral shape of the invariant mass distributions only the systematic errors of the normalization are taken into account as the other systematic errors are assumed to cancel to first order.
For the p+p data a dielectron cocktail was generated using an adapted version of the event generator PYTHIA, see \cite{hades-pp} for details. There are four distinct mass regions: $M_{ee}~$[GeV/c$^{2}$] $<0.15$ (dominated by neutral pion decays), $0.15 <M_{ee}~$[GeV/c$^{2}$] $<0.47$ ($\eta$ Dalitz decay dominated), $0.47 <M_{ee}~$[GeV/c$^{2}$] $<0.7$ (dominated by direct $\rho$ decays and Dalitz decays of baryonic resonances and $\omega$ mesons) and $0.7<M_{ee} ~$[GeV/c$^{2}$] ($\rho$ and $\omega$ dominated) as can be seen from the cocktail. Moreover, around 1 GeV/c$^{2}$ a low statistics $\phi$ signal is visible, which will be discussed in a future publication making use of additional information from its hadronic decay channel.
The underestimation of the dielectron yield in the mass region from $0.47 <M_{ee}~[GeV/c^{2}]<0.7$ points to an insufficient description of the coupling between $\rho$ mesons and baryonic resonances. This coupling will enhance the $e^+e^-$ yield mainly below the $\rho$ pole mass due to kinematical constraints given by the mass distribution of the resonances as well as the ones of the vector mesons \cite{Janus,Janus_new}.
Following vector meson dominance, the coupling of the vector mesons to
baryonic resonances is related to the electromagnetic structure of the
corresponding baryonic transitions. There is then no distinction between the
direct Dalitz decay of baryonic resonances ($N^* \rightarrow N \gamma^*$ ) and the
intermediate coupling to the  rho meson decay ($N^*\rightarrow N\rho\rightarrow N \gamma^*$) and we
will refer to them in the following as "$\rho$-like contribution".\\
\begin{figure}[tb]
   \mbox{\epsfig{figure={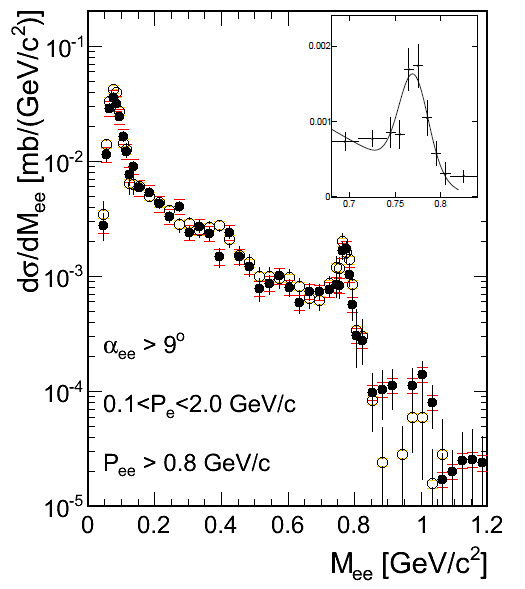}, width=0.49\linewidth}}
   \mbox{\epsfig{figure={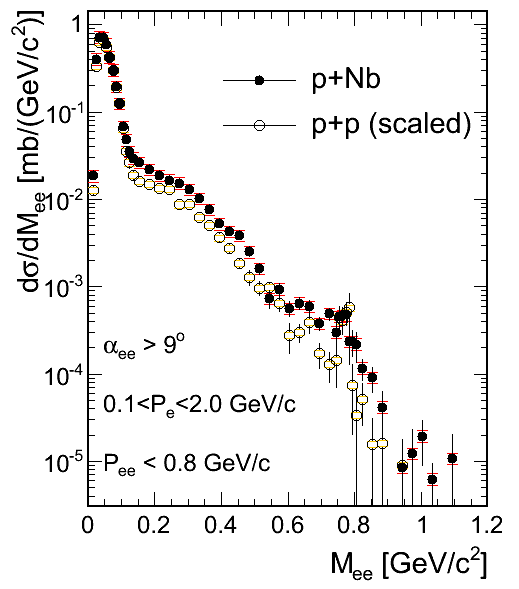}, width=0.49\linewidth}}
   \caption[]{Left: Comparison of the invariant mass spectra for e$^+$e$^-$ pairs with $P_{ee}>0.8$ GeV/c from p+p and p+Nb. The inset shows a linear zoom into the vector meson region together with a fit to the $\omega$ structure for the p+Nb data. Right: For pairs with $P_{ee}<0.8$ GeV/c. The p+p data have been scaled according to a Glauber model.}
\label{pi0}
\end{figure}
\begin{figure}[tb]
   \mbox{\epsfig{figure={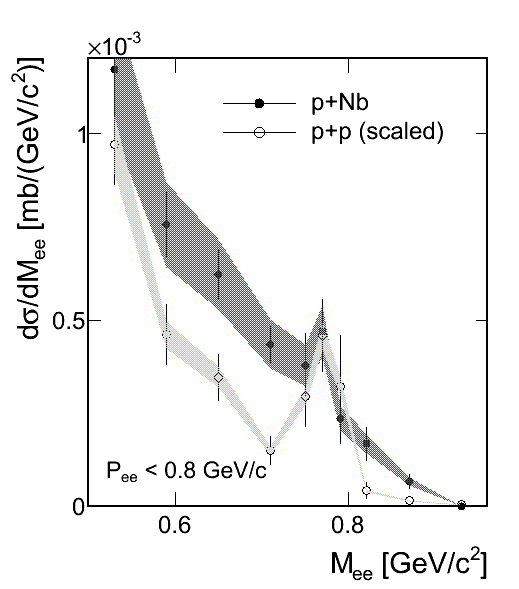}, width=0.49\linewidth}}
   \mbox{\epsfig{figure={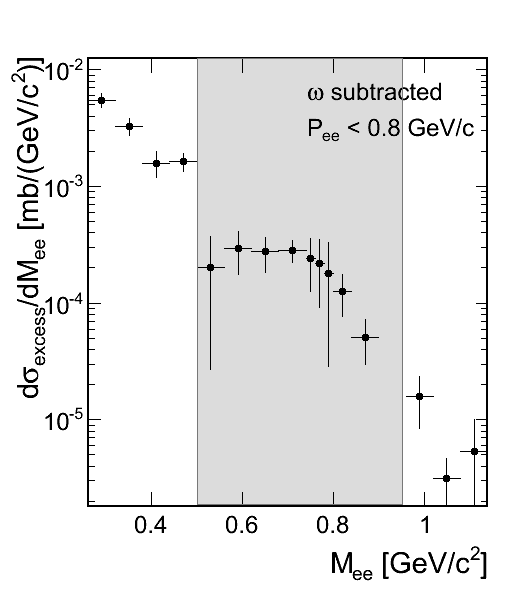}, width=0.49\linewidth}}
   \caption[]{Left: Same as in the right side of Fig.~\ref{pi0} but zoomed into the vector meson region. The shaded bands represent the systematic uncertainties due to the normalization.
   Right: Excess yield in the p+Nb data after subtraction of the scaled p+p reference data (the $\omega$ contribution has been subtracted in both data samples). The grey region corresponds to the invariant mass range plotted in the left picture.
   }
\label{pi2}
\end{figure}
In order to compare the spectral shapes, the p+p data are scaled up according to the nuclear modification factor $R_{pA}$, defined as:
\begin{equation}
R_{pA}=\frac{d\sigma^{pNb}/dp}{d\sigma^{pp}/dp} \times \frac{\langle A_{part}^{pp} \rangle}{\langle A_{part}^{pNb}\rangle} \times \frac{\sigma_{reaction}^{pp}}{\sigma_{reaction}^{pNb}}.
\end{equation}
While $ A_{part}^{pp}= 2$, a value of $ A_{part}^{pNb}=2.8 $ is estimated with the help of a geometrical nuclear overlap model \cite{glauber}. We use $\sigma_{reaction}^{pp}=43.4$ mb from \cite{Bornstein}. This choice of scaling is justified by the agreement of the scaled p+p data with the p+Nb results in the invariant mass region below 150
MeV/$c^2$ (see Fig.~\ref{RpA} below) and the calculations in \cite{Janus_new}.
\newline
A unique feature of the HADES setup is its coverage for low momentum pairs. This allows for the first time to compare the invariant mass distributions for e$^+$e$^-$ pairs with momenta, down to 0.2 GeV/c and larger than 0.8 GeV/c.
The respective contributions are shown in Fig.~\ref{pi0};
for pairs with $P_{ee}>0.8$ GeV/c (left panel) the dielectron yield from p+Nb is slightly lower compared to the scaled p+p data, pointing to absorption of produced mesons inside the nucleus and subsequent particle production in secondary reactions. These second generation particles have then on average smaller momenta and therefore contribute more to the low momentum dielectron sample. The shape of the spectrum is identical to the reference p+p data within errors. Moreover, the width of the $\omega$ peak can be deduced by fitting a Gaussian function to the peak, assuming a smooth background underneath. The corresponding fit, together with a linear zoom into the vector meson region for the p+Nb data, is displayed in the inset of Fig.~\ref{pi0}. Comparing with the p+p data, the results agree within errors ($\Gamma^{\omega_{pole}}_{pp}=16-24$ MeV/$c^2$ and $\Gamma^{\omega_{pole}}_{pNb}=13-19$ MeV/$c^2$), giving no direct hint for broadening of the $\omega$ meson in the nuclear medium.
The situation changes substantially for pairs with $P_{ee}<0.8$ GeV/c. Here one observes a strong e$^+$e$^-$ excess yield below the $\omega$ pole mass, as can be seen in the left panel of Fig.~\ref{pi2}.
Although the e$^+$e$^-$ yield at the $\omega$ pole mass  is not reduced, the underlying smooth distribution is enhanced thus reducing the yield in the peak to almost zero within errors.\\
Due to its large total width, the $\rho$ meson is believed to be the dominating source for radiation from the medium. Therefore we attribute the additional broad contribution to $\rho$-like channels.
The observed decrease of the $\omega$ yield, compared to the p+p indicates a much stronger absorption of $\omega$ mesons than possible feeding from secondary reactions. Unfortunately for low momentum pairs, the extraction of the widths of the $\omega$ peak is hampered by low statistics due to this strong absorption in case of the p+Nb data.\\
This interpretation is in line with previous $\omega$ line shape measurements by the CBELSA-TAPS experiment \cite{Nanova} in the channel $\omega \rightarrow \pi^0 \gamma$ where the $\rho$ decay branch is negligible. The data indicate that, if any in-medium broadening occurs, the change in width of the observed signal is on the percentage level.\\
Following our argumentation, we subtract first, the $\omega$ peak in both data samples and further subtract the scaled p+p dielectron yield from the p+Nb yield. The difference represents the additional e$^+$e$^-$ radiation $\sigma_{excess}$ due to the medium. For the scaled spectra the resulting excess for $P_{ee}<0.8$ GeV/c corresponds to a factor of around $1.5\pm0.3$ of the p+p data in the invariant mass region between 0.3 and 0.7 GeV/c$^2$ and shows an exponential decrease with an additional enhancement directly below the $\rho$ pole mass, i.e. between 0.5 and 0.7 GeV/c$^2$, see right panel of Fig.~\ref{pi2}. Note that this enhancement is exactly at the position where a discrepancy is observed when comparing the p+p data with the PYTHIA calculation (Fig.~\ref{mass}), indicating that both observations might be linked by the same physical process.\\
In order to better understand these observations we compare $R_{pA}$ as a function of the pair momentum $P_{ee}$ in four selected mass regions. Compared to the more abundant particles like pions, the multiplicity of vector mesons is about a factor 50 smaller. Hence the expected feeding of the yield in the vector meson region from pion induced secondary reactions will give a much stronger contribution than the reverse reaction. \newline
\begin{figure}[tb]
   \mbox{\epsfig{figure={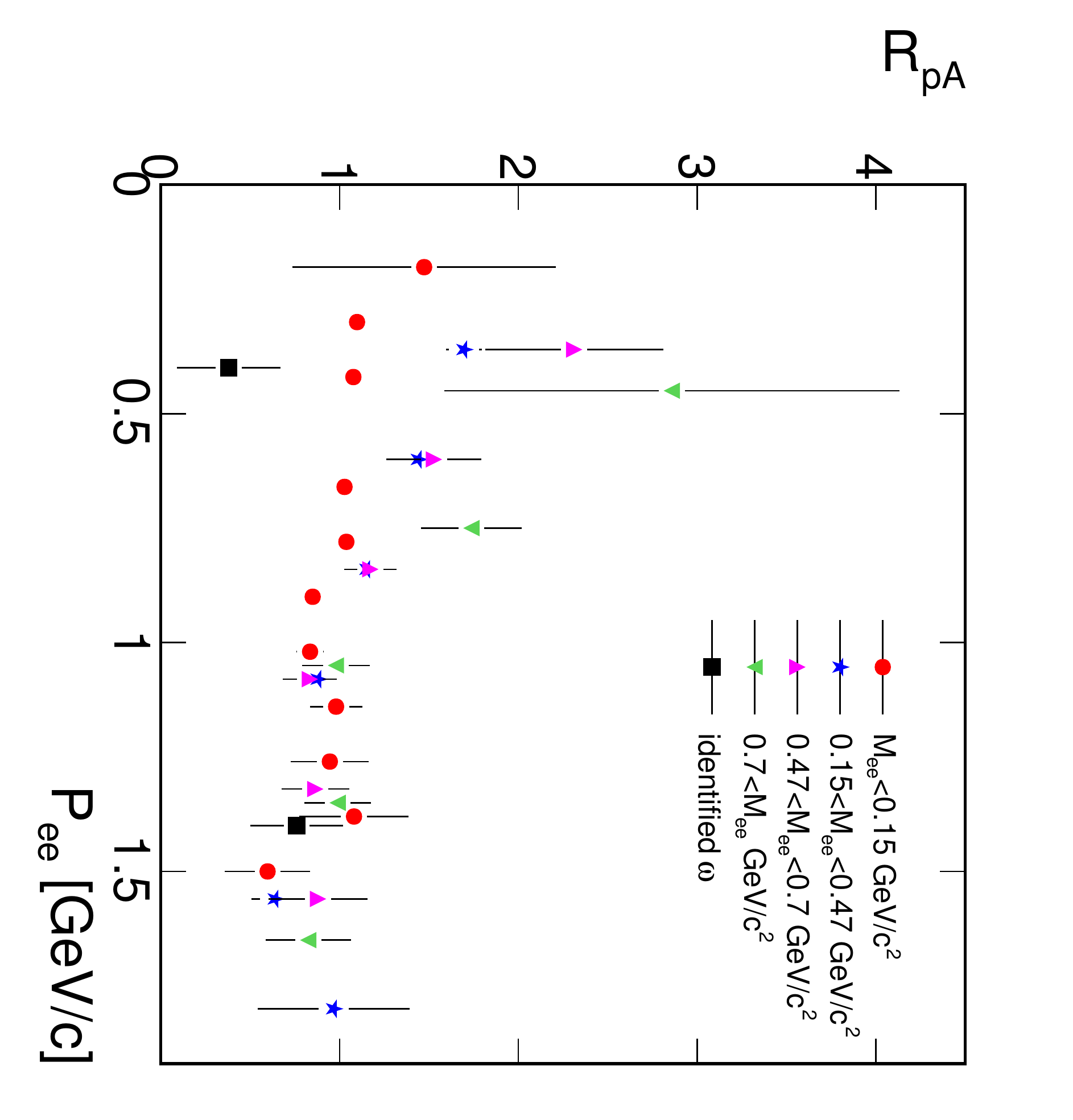}, angle=90, width=0.7\linewidth}}
   \caption[]{Nuclear modification factor $R_{pA}$ as a function of the pair momentum for different invariant mass regions and identified $\omega$. }
\label{RpA}
\end{figure}
In Fig.~\ref{RpA} $R_{pA}$ is shown for four $M_{ee}$ intervals as a function of the $e^{+}e^{-}$ pair momentum. In addition, $R_{pA}$ is also depicted for identified $\omega$ mesons. In absence of nuclear medium effects the value should be unity, hence any deviation indicate additional effects such as, e.g., absorption ($R_{pA} < 1$) or secondary particle production ($R_{pA} > 1$).
While for momenta larger than 1 GeV/c, all $R_{pA}$ values saturate slightly below 1, for smaller momenta $R_{pA}$ rises for all mass regions; the higher the mass, the more pronounced the effect is. At the same time the respective average rapidity $\langle Y_{ee} \rangle$ is shifted below the value corresponding to the free nucleon-nucleon system.
Both observations underline the importance of contributions from secondary reactions for all mass regions.
The $R_{pA}$ of $\omega$ mesons however shows a different trend with values systematically below unity, in strong contrast to the rising factor in the invariant mass region $M_{ee}>0.7$ GeV/$c^2$ (green triangles in Fig.~\ref{RpA}) pointing again to an absorption of $\omega$ mesons which is in line with recent results from the CBELSA-TAPS collaboration \cite{nana}. Therefore a consistent treatment of the spectral shape of vector mesons in the medium together with a correct handling of the additional yield from secondary reactions is needed for an adequate theoretical interpretation.\\
The opposite trends for the $\omega$ and the $\rho$-like contribution might be traced back to their different vacuum properties. As the $\rho$ is already broad in vacuum, any additional broadening inside the medium is not as strongly reflected in a suppression of the partial decay branch, resulting in less suppression due to absorption of the $\rho$ mesons. In this way a natural explanation for our observations would be a suppression of decays outside the medium in favour of additional bulk radiation with a modified spectral distribution.\\
Concerning the nuclear suppression of the $\omega$ signal the $R_{pA}$ values can be directly transformed to a scaling with the nuclear mass number $A$, traditionally used in photon induced reactions, expressed as $\sigma_{\gamma A} \propto \sigma_{\gamma p}\cdot A^{\alpha}$. Here, $\sigma_{\gamma p}$ is the elementary cross section and $\alpha$ is an exponent. For the momentum region $P_{ee}<0.8$ GeV/c, the value of $R_{pA}= 0.38 \pm 0.29$ translates to $\alpha = 0.44 \pm 0.12$, while for $P_{ee}>0.8$ GeV/c and $R_{pA}= 0.76 \pm 0.26$ to $\alpha = 0.67 \pm 0.11$. The latter one agrees with data on proton induced reactions at higher beam energies from the KEK-PS E325 experiment \cite{kek_tabaru} measured also in the $e^{+}e^{-}$ decay channel, where $\alpha = 0.710 \pm 0.021(stat) \pm 0.037(syst)$ is extracted, but for higher momenta.\\
The extracted $\alpha$ values should be compared to photon induced reactions with some caution, due to initial state effects, as discussed in \cite{Bianchi}. In general one expects a weaker scaling in proton induced reactions since the inelastic cross section is large, which leads to shadowing. Hence first chance collisions are located at the surface favoring $A^{2/3}$ scaling over volume scaling \cite{cassing:94}.
Nevertheless, our result agrees with the one obtained by CBELSA-TAPS \cite{cbelsa}, where $\alpha$ values between 0.54-0.74 were found, depending on the meson momentum. Agreement between the two results is also found with respect to the momentum dependence of the $\omega$ peak yield which is explained with cross sections and phase space restrictions in \cite{cbelsa}. Phase space restrictions alone fail however to explain the opposite trend in the $R_{pA}$ for $\omega$ mesons compared to the overall yield in the vector meson mass region, observed in our data.\\
Also restricted to higher momenta are the data of the CLAS experiment \cite{clas2} which reported a stronger absorption compared to both the CBELSA-TAPS result and our data, attributed to $\rho$-$\omega$ interference effects in \cite{clas2}. The latter one should be also present in our data but is not as strongly reflected in the nuclear mass number scaling as observed by CLAS.\\
Although the conditions are extremely different, the qualitative observation of an enhanced $\rho$ contribution and a suppressed $\omega$ yield in radiation from the medium is the same as in dilepton data from heavy-ion collisions in the SPS energy regime by the CERES and NA60 collaborations \cite{Ceres_excess,NA60_rho}.\\
In summary, we have presented the first high statistics measurement of proton-induced dielectron radiation from cold nuclear matter in a kinematic regime, where strong medium effects are expected.
The observed significant difference in the invariant mass spectra in the vector meson mass region can be attributed to a decrease of $\omega$ yield and an enhancement of $\rho$-like contributions, most likely due to secondary particle production.
The importance of secondary collisions is underlined by the enhancement for low momentum dielectrons over the whole invariant mass range in the nuclear modification factor, together with a shift to target rapidity in p+Nb compared to p+p reactions.
The lack of yield in p+p reactions as compared to the PYTHIA calculation, as well as the observed enhancement in p+Nb reactions points to the importance of the coupling of the $\rho$ meson to baryonic resonances which enters in both, $\rho$ meson production (either direct or due to pion-driven secondary reactions) and in-medium modification of the propagating $\rho$.
The observed opposite trend for the $\omega$ can be explained by strong absorption which is another aspect of in-medium modification.\\
We thank our colleagues from theory, in particular Janus Weil for many fruitful discussions.
The collaboration gratefully acknowledges the support
by CNRS/IN2P3 and IPN
Orsay (France), LIP Coimbra, Coimbra (Portugal): PTDC/FIS/113339/2009,
SIP JUC Cracow, Cracow (Poland): N N202 286038 28-NN202198639, Helmholtz-Zentrum Dresden-Rossendorf (HZDR),
Dresden (Germany): BMBF 06DR9059D, TU Munchen,
Garching (Germany) MLLMuenchenDFG EClust: 153VHNG-
330, BMBF 06MT9156 TP5 TP6, GSI TMKrue 1012,
NPI AS CR, GSI TMFABI 1012, Rez, Rez (Czech Republic):
MSMT LC07050 GAASCR IAA100480803, USC - S.
de Compostela, Santiago de Compostela (Spain): CPAN:CSD2007-
00042, Helmholtz alliance HA216/EMMI.

\end{document}